\documentclass[pra,aps,onecolumn, nofootinbib]{revtex4}
\pdfoutput=1
\usepackage[T1]{fontenc}
\usepackage{amssymb,bm}
\usepackage{amsthm} 
\usepackage{amsmath} 
\usepackage{graphicx}

\def\be{\begin{equation}}
\def\ee{\end{equation}}

\newcommand{\ket}[1]{\left\vert{#1}\right\rangle}

\theoremstyle{definition} 

\theoremstyle{theorem} 

\theoremstyle{theorem}

\theoremstyle{definition}

\begin{document}

\title{Private quantum computation: An introduction to blind quantum computing and related protocols}  
\author{Joseph F. Fitzsimons}
\email{joseph_fitzsimons@sutd.edu.sg}
\affiliation{Singapore University of Technology and Design, 8 Somapah Road, Singapore 487372}
\affiliation{Centre for Quantum Technologies, National University of Singapore, 3 Science Drive 2, Singapore 117543}

\begin{abstract}
Quantum technologies hold the promise of not only faster algorithmic processing of data, via quantum computation, but also of more secure communications, in the form of quantum cryptography. In recent years, a number of protocols have emerged which seek to marry these concepts for the purpose of securing computation rather than communication. These protocols address the task of securely delegating quantum computation to an untrusted device while maintaining the privacy, and in some instances the integrity, of the computation. We present a review of the progress to date in this emerging area.
\end{abstract}

\date{\today}
\maketitle

\section{Introduction} 

For almost as long as programmable computers have existed, there has been a strong motivation for users to run calculations on hardware that they do not personally control. Initially, this was due to the high cost of such devices coupled with the need for specialised facilities to house them. Universities, government agencies and large corporations housed computers in central locations where they ran jobs for their users in batches. Over time computers have become ubiquitous, but demand for centralised resources has not abated. Even today, the use of delegated computation is widespread, in the form of cloud computing.

While we do not yet know how the field of quantum computing will develop, it seems reasonable to speculate that it will follow a similar path. Indeed this speculation is somewhat born out by recent efforts to provide access to rudimentary quantum processors over the Internet \cite{steffen2016progress}. Today we are in a far better position to enable remote access to quantum computers than was possible with early conventional computers, due to the existence of high speed global communications networks, and the ubiquity of classical processors. Furthermore, the discovery of quantum key distribution protocols \cite{bennett1984quantum,ekert1991quantum} has provided the impetus to develop quantum communication over existing optical fibre networks \cite{patel2012coexistence}. These factors only serve to increase the scope for early adoption of delegated quantum computation.

While the option of delegating calculations to remote systems may have strong practical and economic motivation, it opens a myriad of security concerns. In particular, if the computation is performed on untrusted hardware, then this opens the possibility that either the privacy or the integrity of the computation may be compromised. Encryption can be used to hide communication between the client and the server from eavesdroppers, while authentication codes can be used to detect any attempt to modify these messages \cite{ferguson2011cryptography}. However, such techniques do nothing to counteract the threat posed by a compromised or malicious server. Ideally, to overcome these concerns, one would want a way to delegate tasks to a remote server while ensuring privacy, even from the server executing them, and to ensure the correctness of the result. 

In recent years, a number of protocols have emerged which seek to tackle the privacy issues raised by delegated quantum computation. Going under the broad heading of blind quantum computation (BQC), these provide a way for a client to execute a quantum computation using one or more remote quantum servers while keeping the structure of the computation hidden. While the goal of BQC protocols is to ensure only the privacy of the computation, many also allow for verification of the computation being performed, by embedding hidden tests within the computation.

To date blind quantum computation has been considered in a wide range of settings, with varying requirements on the capability of the client and the server or servers. Ultimately, the most desirable setting would be a verifiable blind quantum computation protocol which could be performed between a client without any quantum capabilities and a single quantum server. Unfortunately, progress on such a protocol has proved slow. Part of the difficulty is that the server could retain a complete transcript of the communication during the protocol, allowing them to rerun their side of the process many times. Indeed, a no-go result from Morimae and Koshiba ruled out a wide class of potential protocols \cite{morimae2014impossibility}. Furthermore, results from classical secure computing create a link between blind computing and computational complexity \cite{abadi1987hiding}. The existence of a sufficiently secure blind computation protocol with a purely classical client and a single quantum server capable of implementing arbitrary quantum computations would create a link between the questions of whether BQP contains NP and whether the polynomial hierarchy collapses \cite{dunjko2016blind}. As a result of these hurdles, it is only very recently that mechanisms which may allow for such functionality have begun to emerge \cite{mantri2016flow}.

As a result, progress on blind computation has come from considering settings which relax these restrictions somewhat. There are several ways in which this can be done, which can broadly be divided into two categories: Settings which relax the requirement that the client be purely classical, and settings which allow for multiple non-communicating quantum servers. Settings considered in the first category augment the client with some quantum capability which is insufficient for quantum computation unaided. The motivation behind this approach is that it may be plausible to allow the client to prepare or measure single qubit states \cite{broadbent2010measurement,morimae2013blind}, or perhaps to have a small quantum processor of their own \cite{aharonov2010proceedings}. On the other hand, the second category maintains a classical client, but allows that client to interact with multiple servers \cite{reichardt2013classical}. In all known protocols of this type, the privacy of the computation is only maintained provided that the servers do not communicate. While it is in principle possible to achieve this for short periods by using position to enforce spacelike separation between servers during the protocol, it may be difficult to guarantee blindness indefinitely in any realistic scenario.

\section{Security \label{sec:security}}

Before diving into the protocols which have been proposed to accomplish blind quantum computation, it is necessary to clarify what precisely this statement means at a more formal level. Just as many configurations of client and server have been considered in the literature, a variety of security definitions have also been proposed. At an intuitive level, security definitions for this problem seek to capture, at a minimum, the idea that a malicious server or servers should be unable to distinguish between possible computations chosen by the client, based on the information they receive during the protocol. Certain information cannot be hidden. The server will always know the resources they committed to the computation and so can always determine an upper bound on depth and width of the quantum circuit corresponding to the client's chosen computation. However, it is in principle possible to hide all other aspects of the computation and to pad the circuit so that the circuit dimensions reveal little information about the hidden computation. In formulating a security definition for blind computation, it is hence important to account for this leakage of information. In this section, we will focus primarily on security definitions that are information theoretic in nature, avoiding the necessity for assumptions on the computational power of the adversary which are characteristic of much of modern classical cryptography.

The term {\em blind quantum computation} was first coined by Arrighi and Salvail \cite{arrighi2006blind} to describe the encryption of instances of certain problems so that they could be delegated while preserving privacy, building upon similar ideas introduced in the classical regime by Feigenbaum \cite{feigenbaum1985encrypting}. The first protocol and security definition for securely delegating arbitrary quantum computations, termed {\em universal} blind quantum computation, was introduced in \cite{broadbent2009universal}. There blindness was defined in the following way. Given a description of a client's computation $x$, a protocol was said to be {\em blind while leaking at most $L(x)$} if the distribution of classical and quantum information received by the server was fully determined by $L(x)$. The purpose of the function $L(x)$ is to capture information unavoidably leaked, which may differ from protocol to protocol but is most often taken to be the circuit dimensions of the delegated computation. In effect, this definition demands that the information received by the server not depend on any aspect of the client's chosen computation other than what is captured by $L(x)$. Much of the work on blind computation to date has implicitly or explicitly used some form of this definition. A notable exception to this trend was presented in \cite{giovannetti2013efficient} which made use of a weakened definition, requiring only that attempts to violate the privacy of the computation be detected with high probability, in order to optimise communications overhead. 

While the previous definition captures the spirit of blind quantum computation, it has certain undesirable features. In particular it does not fully determine how a protocol satisfying that definition will behave as part of a larger system. For example, if the server were to deviate from the protocol, they would still only learn at most $L(x)$. However, if subsequent to the conclusion of the protocol, they were to learn something about the resulting state received by the client, it might be possible for them to infer additional details of the computation. To this end, stronger security definitions for delegated computation have been proposed which seek to fully define the behaviour of BQC protocols \cite{dunjko2014composable} using the abstract cryptography framework of Maurer and Renner \cite{maurer2011abstract}. 

\begin{figure}[!h]
\includegraphics[width=\columnwidth]{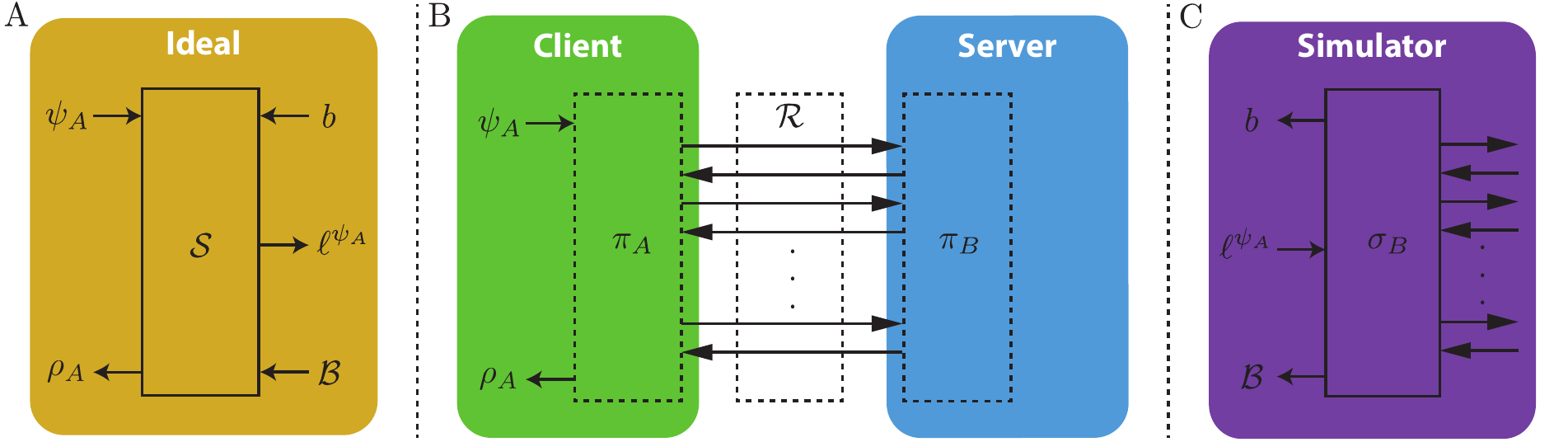}
\caption{A) A schematic of the ideal resource for blind computation. Here the convention is used that the client's interfaces are on the left, while the server's interfaces are on the right. B) A representation of how the two protocols $\pi_A$ and $\pi_B$ are combined with a communications resource $\mathcal{R}$ in order to implement a channel mapping $\psi_A$ to $\rho_A$. C) A schematic of a simulator which can be appended to the right side of ideal resource so that the interfaces of the compound object match those of $\pi_A \mathcal{R}$.\label{fig:security}}
\end{figure}

This is achieved through the use of the notion of an ideal resource, which fully describes the functionality of blind computation without making reference to any particular protocol. This can be thought of as a black box process with several interfaces through which the client and server can interact with it, as shown in Figure \ref{fig:security}A. Several of these interfaces allow for the passing of quantum information, which may in general be in a non-separable state. Hence the labels on each interface refer to quantum systems rather than to the corresponding states directly, such that for example $\psi_A$ and $\psi_B$ are two subsystems of a compound system $\psi_{AB}$. The system $\psi_A$ represents the description of the client's computation together with any included quantum input, such that the output of the chosen computation is given by $\mathcal{U}(\psi_A)$ for a fixed unitary operator $\mathcal{U}$ known to both the client and the server. This use of a quantum state to represent the description of the computation as well as the input can be thought of as providing input for a programmable device which includes both the programme and the input data, and can be done without loss of generality. The system $\rho_A$ represents the output obtained by the client, while $\ell^{\psi_A}$ represents the leaked information obtained by the server based on the client's computation. In general these can all be considered to be quantum states, though for specific use cases they may be classical. A single bit, $b$, is used to indicate whether the server chooses to deviate from the protocol. The specification of $\mathcal{B}$ and the relationship between the output $\rho_A$ and the other inputs depends on whether or not verification is included. In the case where only blindness is considered, the description of the ideal resource, denoted $\mathcal{S}^\text{blind}$ is completed by choosing $\mathcal{B} = (\mathcal{E},\psi_B)$, with $\mathcal{E}$ being a completely positive trace-preserving map (capturing the server's interference with the computation) and $\psi_B$ being a quantum system, and fixing $\rho_A = \mathcal{U}(\psi_A)$ when $b=0$ and $\rho_A = \mathcal{E}(\psi_{AB})$ when $b=1$, where $\psi_{AB}$ is the joint system composed of subsystems $\psi_A$ and $\psi_B$. In the case where verification is included, the ideal resource $\mathcal{S}^\text{blind}_\text{verif}$ is completed by choosing $\mathcal{B}$ to be a single bit $c$. If $c=0$ then $\rho_A = \mathcal{U}(\psi_A)$, otherwise when $c=1$ the output gives some fixed error state $\rho_A = |\text{err}\rangle\langle\text{err}|$ orthogonal to the usual output space. Thus $c$ captures the possibility of the server causing an error in the computation, but requires that it be perfectly detectable by the client.

With the ideal resources specified, security definitions can then be given which relate a concrete protocol to the ideal resource. A concrete BQC protocol is composed of a pair of protocols, $\pi_A$ for the client and $\pi_B$ for the server, which interact via a communications channel indicated by $\mathcal{R}$, as illustrated in Figure \ref{fig:security}B. In this framework, a BQC protocol should satisfy two conditions in order to be considered secure, correctness and blindness. Correctness captures the notion that the output of the proposed protocol actually matches the behaviour of the ideal resource when the server is behaving honestly. We can consider the composition of protocols and resources using $\alpha\beta$ to denote the composition of two such objects where the right side interfaces of $\alpha$ match and are connected to the left hand interfaces of $\beta$. A protocol is then said to be $\epsilon$-correct if the quantum channel implemented by $\pi_A\mathcal{R}\pi_B$ is $\epsilon$-close in the diamond norm to the channel given by $\mathcal{S} \bot_B$, where $\bot_B$ denotes that the server is restricted to behaving honestly and so obstructs his interfaces (setting $b=0$ and $c=0$). Here $\mathcal{S}$ is taken to be either $\mathcal{S}^\text{blind}$ or $\mathcal{S}^\text{blind}_\text{verif}$ depending on whether verifiability is included. The blindness condition captures the notion that the server should not be able to learn more than if they were interacting with the ideal functionality, but is stronger than the definition considered earlier. A protocol is $\epsilon$-blind if $\pi_A \mathcal{R}$ is $\epsilon$-close to $\mathcal{S}\sigma_B$ for some {\em simulator} $\sigma_B$ in the appropriate norm. This is stronger than demanding that the server learn no more than they can from the ideal resource, since it implies that any party having access both to the server and client interfaces of the compound objects cannot distinguish $\pi_A \mathcal{R}$ from $\mathcal{S}\sigma_B$. This is an important feature, since it means that in analysis of larger systems which makes use of blind computation as a component, the concrete BQC protocol can be replaced by an ideal functionality, provided $\epsilon$ is sufficiently small. For a more thorough treatment of composable security definitions for blind computation, curious readers are referred to \cite{dunjko2013composable}.

\begin{figure}[!h]
\includegraphics[]{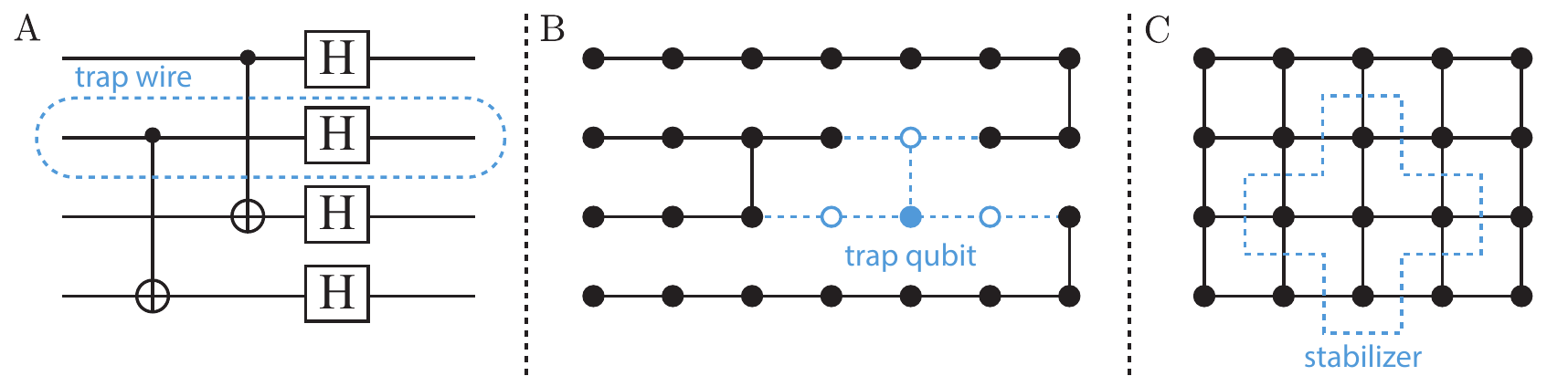}
\caption{Approaches to verifying blind quantum computation. A) The approach taken to verification in \cite{broadbent2009universal} and \cite{broadbent2015verify}. The client chooses the computation such with some fixed probability each logical qubit may be a ``trap'', for which the outcome can be easily computed by the client and used to detect any deviation from the protocol by the server. B) In \cite{fitzsimons2012unconditionally} individual physical qubits, rather than logical qubits, are used as traps. By hiding $Z$-basis measurements on the neighbouring qubits the client can surreptitiously disentangle a chosen qubit from the rest of the resource state. The net result is that the outcome for the measurement of the trap qubit in an honest run of the protocol is known in advance by the client, and hence can be used to verify that the server has not introduced an error into the computation. C) In the setting where the client performs adaptive measurements on a fixed resource state, performing measurements on the received qubits to ensure that they satisfy the same stabiliser relations as the ideal resource offers an alternative mechanism to ensure that the server behaves honestly. This approach was initially proposed in \cite{morimae2014verification}. \label{fig:traps}}
\end{figure}

The above discussion focuses on blindness or the combination of blindness and verifiability, rather than on verifiability alone. A much wider variety of non-equivalent definitions has been considered for this latter property, and a review of all such definitions is beyond the scope of the current manuscript. There does, however, appear to be a deep link between notions of blindness and verification. Several protocols designed primarily to verify quantum computation have turned out to yield BQC protocols with minimal or no changes \cite{aharonov2010proceedings,reichardt2012classical}, while the notion of embedding hidden traps within a computation, as shown in Figure \ref{fig:traps}, has been used to make several BQC protocols verifiable \cite{fitzsimons2012unconditionally,morimae2014verification}. Recently, however, several verification schemes have emerged which do not seem to immediately give rise to blind computation protocols \cite{fitzsimons2015post,morimae2016post,natarajan2016robust}, and so it remains an open question as to whether or not these are truly independent properties.

\section{BQC with semi-classical clients}

\subsection{Restricted quantum computation \label{sec:computing}}

\begin{figure}[!h]
\includegraphics[width=\columnwidth]{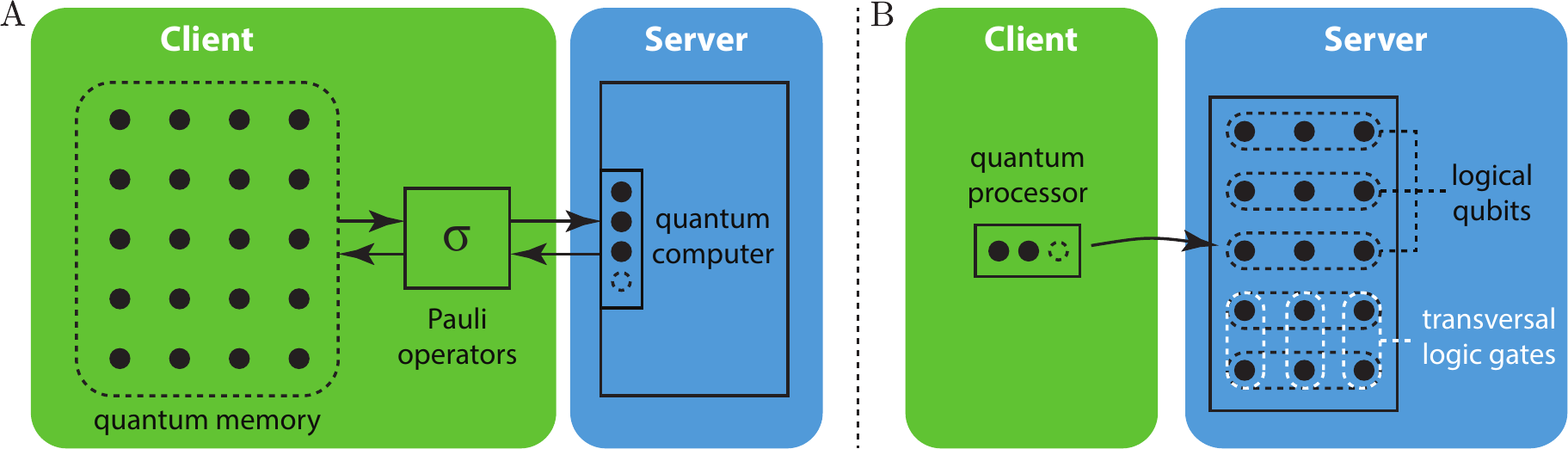}
\caption{A) An illustration of the BQC setting considered by Childs \cite{childs2005secure}. In this scenario, the client has a large quantum memory together with the ability to perform Pauli operations on qubits and to transmit them to the server. However they lack the ability to perform other gates, such as Toffoli or Hadamard gates. B) An illustration of the setting considered by Aharonov, Ben-Or and Eban \cite{aharonov2010proceedings}, in which the client has access to a quantum computer capable of performing arbitrary operations on a constant number of qubits.\label{fig:CABE}}
\end{figure}
The first blind quantum computation protocol is widely attributed to Childs \cite{childs2005secure}, who introduced an interactive protocol which allowed a user with restricted quantum capabilities to perform universal quantum computation with the aid of a second party possessing a universal quantum computer, while keeping the specifics of the computation hidden. In the scenario originally considered, the client had a large quantum memory with the ability to rearrange qubits and to perform Pauli operations, but lacked the ability to perform non-Pauli gates and to perform measurements. The ability to perform Pauli gates allowed the client to encrypt qubits using a quantum one-time pad (i.e.~applying a random Pauli to each qubit) before sending them to the server. The client selects a qubit or qubits to perform an operation on, applies a random quantum one-time pad, and transmits them to the server. The server then applies the desired operation before returning the qubits to the client. When encoded in this way, measurements made by the server revealed no information about the state of the encoded qubit, but can be decoded by the client using the one-time pad key. Furthermore, Clifford group gates can be applied by the server directly onto the encrypted state provided that the client update their encryption key. This is due to the fact that given a Clifford group operator $C$, a multi-qubit Pauli operator $\sigma$ and and quantum state $\ket{\psi}$, $C\sigma\ket{\psi} = \sigma' C \ket{\psi}$ where $\sigma' = C\sigma C^\dagger$. The procedure for implementing non-Clifford group gates is a little more involved. Consider the result of the server applying the gate $T = |0\rangle\langle0| + e^{i \frac{\pi}{4}}|1\rangle\langle 1|$ to a state encrypted with a quantum one-time pad given by the Pauli operator $\sigma$. If $\sigma$ commutes with $T$, then trivially $T \sigma \ket{\psi} = \sigma T\ket{\psi}$ and the gate has been successfully applied on the encrypted state. If however $\sigma$ does not commute with $T$, then it is because $\sigma$ contains either an $X$ or $Y$ term corresponding to the qubit on which $T$ is being applied. In this case $T \sigma \ket{\psi} = \sigma T^\dagger \ket{\psi}$. While this is not the desired result, it is possible to make a correction by applying the gate $S=T^2$, which is in the Clifford group and hence can be applied deterministically by the server using the previous procedure. One caveat is that if the server is only requested to apply an $S$ gate following a $T$ gate if $\sigma$ does not commute with $T$ on the chosen qubit, then information about the encoded state is revealed. In order to avoid this scenario, following each $T$ gate the server must always be requested to perform an $S$ gate. In the case where $S$ is unnecessary, the client simply chooses an ancillary qubit for it to be performed on. Since the qubit is returned to the client between each step, and the one-time pad renewed, there is no mechanism for the server to distinguish between the two cases. As these operations together are sufficient for approximately-universal quantum computation, the client can thus make use of the server to implement an arbitrary quantum computation without revealing their state. This can be extended to hide the full computation by requiring that the server always implement a fixed order of gates (say Hadamard, CNOT, $T$, $S$) where the client simply choses to have unwanted gates applied to an ancillary pair of qubits. 

A notable feature of this protocol is that the quantum resources required of the client depends on the computation being performed. Furthermore, while Childs discussed the issue of verification, noting that the computation could be verified by a classical client if it resulted in a witness for an instance of a problem in NP and speculating on the use of tomography on a subset of gates as a mechanism for keeping a memoryless server honest, the protocol did not provide a general mechanism for verifying the correctness of general computation. Subsequent work has sought to both reduce the requirements on the client and to ensure verifiability of the computation. Specifically, an alternate approach taken by Aharanov, Ben-Or and Eban in the context of developing interactive proofs for BQP, which shifts the memory requirement to the server \cite{aharonov2010proceedings}. In the ABE protocol, the client is used to prepare the initial state for the computation, encoded using a polynomial code introduced in \cite{ben2006secure} which amounts to a quantum authentication scheme \cite{barnum2002authentication}. This can be done logical qubit by logical qubit, and so the size of the client's device need not scale as a function of the number of qubits in their chosen computation, but can be as low as 3 qubits. The algebraic structure of the encoding allows for the server to implement Clifford gates transversally, provided that the client update their encryption key, and so by making use of gate teleportation using magic states prepared by the client it is possible to implement an approximately universal gate set. Just as in the case of Childs' protocol, the encoding hides the client's input state only. However, by making use of a fixed programmable unitary, similar to that used in the composable security definitions discussed earlier, this can be leveraged into a BQC protocol, hiding not only the input but also the intended computation. Furthermore, the use of an authentication code ensures that the computation can be verified with constant probability of error.

\subsection{State preparation}

While the approaches described in the previous section do allow a client with limited quantum capabilities to harness the power of a larger quantum computer, they require reasonably sophisticated apparatus on the client's side. As we shall see, it is possible to further reduce the complexity of the client's device. The universal blind quantum computing protocol (UBQC) introduced in \cite{broadbent2009universal} manages to hide arbitrary quantum circuits with a client only capable of preparing certain single qubit quantum states. Rather than directly implementing a computation as a series of gates applied to a fixed register of qubits, the UBQC protocol implements the desired circuit as a measurement-based computation.

\begin{figure}[!h]
\includegraphics[]{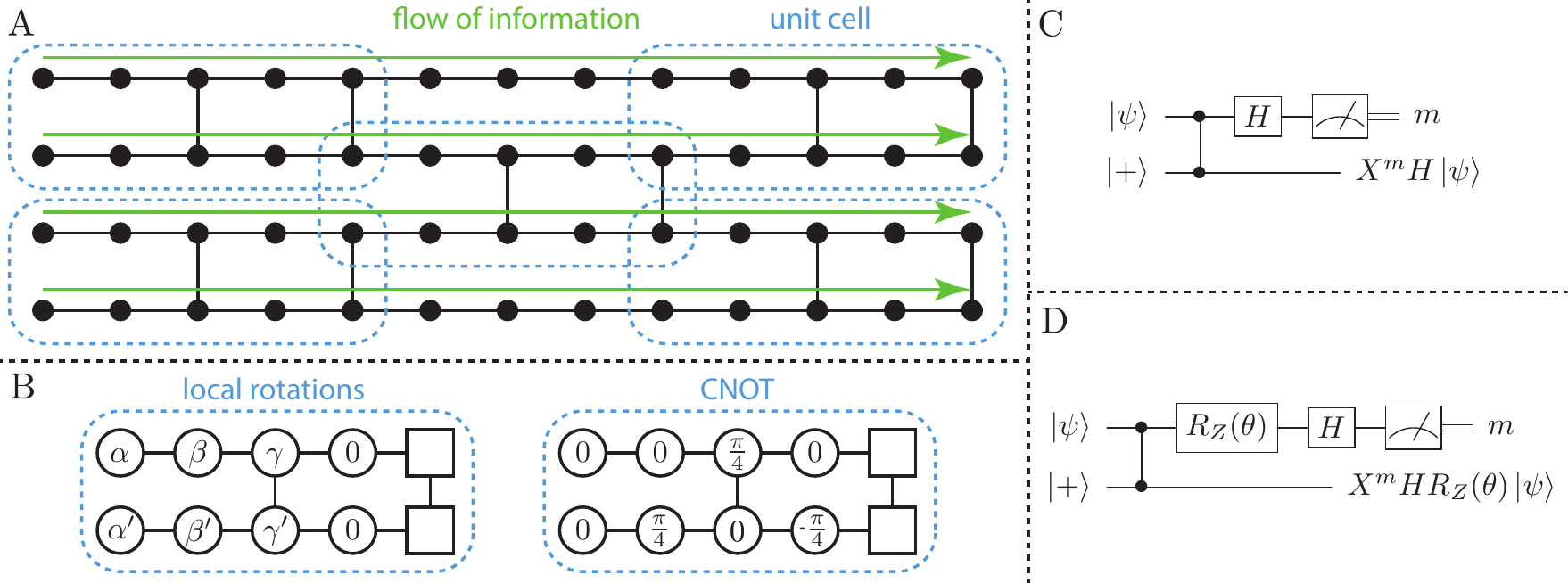}
\caption{The structure of measurement-based computation using the brickwork state as the resource. A) The brickwork state underlying the UBQC protocol \cite{broadbent2009universal}. The same resource state has previously been considered in \cite{childs2005unified}. Logical qubits propagate along horizontal chains of vertices from right to left. These can be thought of as corresponding to wires in a quantum circuit diagram, with each vertex corresponding to the application of a single qubit gate, and vertical edges corresponding to controlled-phase gates between neighbouring qubits. The graph of the resource state is constructed from a regular tiled unit cell. B) Possible choices for measurements within the unit cell giving rise to arbitrary single qubit rotations on each logical qubit or to a CNOT gate. C) A two qubit teleportation circuit which can be used to understand the propagation of logical qubits along the horizontal chains indicated in A when qubits are measured in the $X$ basis. D) A modified version of the two qubit teleportation protocol which can be used to understand the effect of measurements in arbitrary bases in the $XY$-plane when propagating logical qubits through the graph, as discussed in \cite{nielsen2006cluster}.\label{fig:brickwork}}
\end{figure}

In order to understand how the UBQC protocol works, it is first necessary to understand how a computation can be expressed in the measurement-based model. In measurement-based quantum computation (MBQC), a quantum computation is expressed as a sequence of single-qubit measurements to be performed on a fixed resource state \cite{briegel2009measurement}. These resource states are known as graph states, due to their correspondence to simple graphs. The resource state corresponding to a particular graph $G$ can be determined as the output of a circuit specified as follows. For each vertex in $G$, prepare a qubit in the state $\ket{+} = \frac{1}{\sqrt{2}}\left(\ket{0} + \ket{1}\right)$. Then, for each edge $e$ in $G$, apply a controlled-phase gate between the qubits corresponding to the vertices joined by $e$. Since the controlled-phase gates necessarily commute, the state produced by the circuit is independent of the order of these operations. While the original resource states studied in the context of MBQC were cluster states \cite{raussendorf2001one,raussendorf2003measurement}, the graph for which corresponds to a regular square lattice, the UBQC protocol made use of a different resource state known as a brickwork state, illustrated in Figure \ref{fig:brickwork}. The reason for this was that the initial insight underlying the UBQC protocol allowed the hiding of only measurements which projected onto states in the $XY$-plane of the Bloch sphere. Until very recently cluster states were not known to be universal without the addition of $Z$-basis measurements, and so a resource state was constructed which required only $XY$-plane measurements. Recent results proving the universality of cluster states with only $XY$-plane measurements imply that the UBQC protocol could be trivially modified to use such states \cite{mantri2016universality}. For simplicity, in discussing how MBQC implements a computation, we will consider only the case of the brickwork state. Readers interested in a more thorough introduction to this model of computation are referred to \cite{browneone}.

In a measurement-based computation performed on a brickwork state, the planar nature of the underlying graph has a natural interpretation in terms of the circuit model. Each row of vertices, together with the edges connecting them, corresponds to a single logical qubit which is propagated from left to right by a sequence of measurements, as shown in Figure \ref{fig:brickwork}A. The initial state of each such logical qubit is $\ket{+}$ and initially the state can be thought of as residing at the leftmost vertex in the chain. Each measurement in a basis $B_{\theta} = \cos(2\theta) X + \sin(2\theta) Y$ has the effect of propagating the logical qubit on vertex to the right and applying the operator $X^m H R_Z(\theta)$, where $R_Z(\theta) = \cos\theta I + \sin\theta Z$ and $m \in \{0,1\}$ is the outcome of the measurement. This is a consequence of a rotated version of the two-qubit teleportation protocol illustrated in Figures \ref{fig:brickwork}C and \ref{fig:brickwork}D \cite{childs2005unified,nielsen2006cluster}. The presence of the teleportation byproduct $X^m$ means that subsequent measurement angles need to be adapted to negate this byproduct in order to achieve deterministic computation. Thus the measurement bases will in general depend on the outcomes of previous measurements as well as the intended logic gate they correspond to, a dependency formalised for general graphs by the notion of flow \cite{danos2006determinism}. The only remaining element unaccounted for, then, are vertical edges in the graph. These can be seen as controlled-phase gates between logical qubits which occur when both logical qubits have been propagated onto the vertices linked by a vertical edge. Measurement of the rightmost qubit in each row corresponds to a measurement of the final state of the computation. Taken together, these two elements can be combined to perform more common universal gate sets such as arbitrary local unitary operations and CNOTs as shown in Figure \ref{fig:brickwork}B.

\begin{figure}[!h]
\includegraphics[]{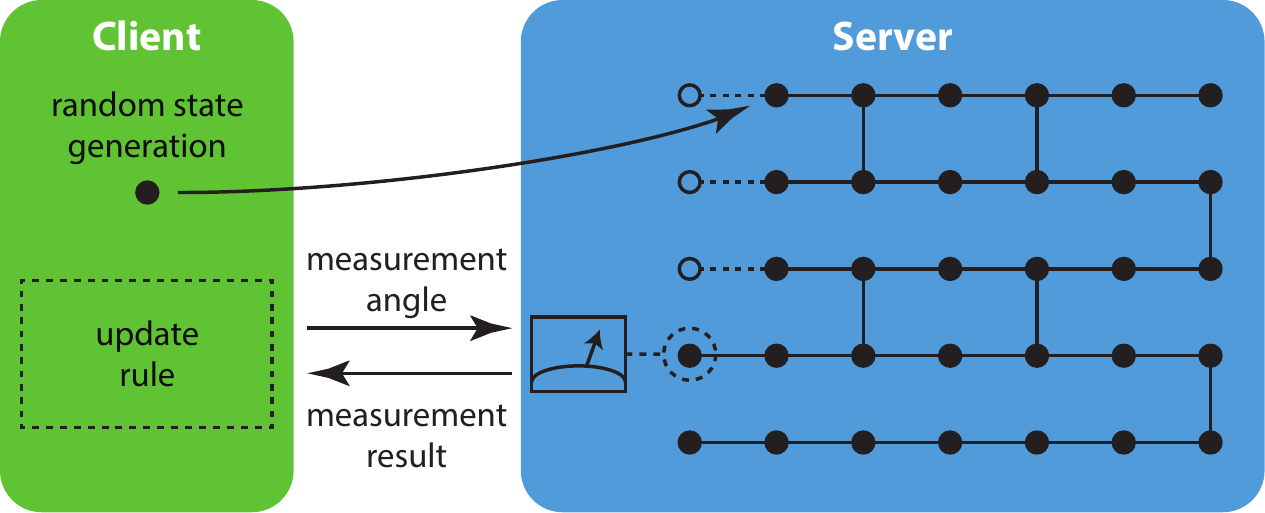}
\caption{A depiction of the Universal Blind Quantum Computation (UBQC) protocol introduced in \cite{broadbent2009universal}. The client prepares random single qubit states which are then sent to the server to be entangled. For each qubit in turn, the client computes a measurement angle which is sent to the server, which returns the result to the client for inclusion in the calculation of subsequent measurement angles. \label{fig:ubqc}}
\end{figure}

UBQC was the first protocol to take advantage of measurement-based computation in the context of blind computation. It achieves blindness by effectively hiding the measurement bases for a measurement-based computation on a fixed resource state. The UBQC protocol, depicted in Figure \ref{fig:ubqc}, proceeds as follows. The client first encodes their computation as a sequence of measurements at angles $\phi_i$, chosen from the set $A = \{0,\frac{\pi}{4},\frac{2\pi}{4},\ldots, \frac{7\pi}{4}\}$, in the $XY$-plane on a brickwork state of dimensions $N \times M$, proportional to the dimensions of the quantum circuit\footnote{The circuit is restricted to nearest neighbour interactions, which can be done without loss of generality.} corresponding to the desired computation. The client then prepares and sends $NM$ single qubit states $\ket{\psi_i} = \frac{1}{\sqrt{2}}\left(\ket{0} + (-1)^{r_i} e^{i \theta} \ket{1}\right)$, where $r_i$ and $\theta_i$ are chosen uniformly at random from the sets $\{0,1\}$ and $A$ respectively, and transmits them to the server\footnote{Subsequent work from Dunjko and Kashefi has shown that some level of blindness can be maintained with the client preparing qubits in only two possible states \cite{dunjko2016blind}}. The server arranges these in a two dimensional grid of $N \times M$ qubits and entangled them with controlled-phase gates according to the brickwork graph. The computation then proceeds in rounds in which the qubits are measured sequentially, in a fixed order (going from top to bottom, left to right). In the $i$th round, the client sends a measurement angle $\delta_i = \phi_i' - \theta_i$ to the server, who measures qubit $i$ in this basis and returns the resulting outcome $b_i$ to the client. Here $\phi'_i$ denotes updated measurement angle for qubit $i$ adapted from $\phi$ to account for previous teleportation byproducts. The client then decodes this outcome to obtain a new bit $m_i = b_i \oplus r_i$, which they take to be the true outcome of the measurement. This entire procedure can be seen to be equivalent to the original measurement-based computation chosen by the client since rotations about the $Z$ axis commute with the entangling operations and hence the $\theta_i$ terms included in the state preparation and measurement angle cancel. Furthermore, the effect of $r_i$ is equivalent to applying a $Z$ operation to the initial state, which commutes with the entangling operations but anti-commutes with the measurement operator, and hence results in a bit flip on the outcome of the measurement result $b_i$, which is undone in the computation of $m_i$. It is easy to see that the UBQC protocol satisfies the conditions of the first blindness definition introduced in Section \ref{sec:security} by noting that only $\ket{\psi_i}$ depends on $r_i$, and as these values are random, and a priori unknown to the server, the density matrix for the state received by the server is always maximally mixed, and so fixed and independent of $\theta_i$. Thus only $\delta_i$ is dependent on $\theta_i$, which is chosen uniformly at random from the same set as $\phi_i$ and is a priori unknown to the server, and hence it too is uniformly random and independent of $\phi_i$ or $\phi_i'$. As such, the distribution of messages sent to the server, when averaged over choices of the random variables $\{r_1,\ldots,r_{NM}\}$ and $\{\theta_1,\ldots,\theta_{NM}\}$ is fixed as the maximally mixed distribution. Blindness under this first definition was proved in \cite{broadbent2009universal}, while security under the stronger composable definition proved in \cite{dunjko2013composable}. It is worth noting that while the resource state requires $NM$ qubits, due to the commutation of operations involving non-neighbouring qubits, not every qubit needs to be present in the initial state. Indeed, by postponing each controlled-phase operation as long as possible, it is possible to implement this protocol using only $N+1$ qubits.

While the original paper introduced a mechanism for verifying the delegated computation based on the use of ancillary {\em trap} qubits randomly interspersed with the target computation (see Figure \ref{fig:traps}A), the analysis of verification given in \cite{broadbent2009universal} was incomplete. Subsequently a modified verification technique was proposed based on using physical qubits rather than logical qubits as traps (see Figure \ref{fig:traps}B) and proven to suppress arbitrary deviations from the protocol by the server \cite{fitzsimons2012unconditionally}. The modified BQC protocol also introduced new functionality, making it possible to incorporate hidden $Z$-basis measurements within a measurement pattern, and allowing for the entanglement graph to be hidden. This modified protocol also satisfies the composable security definitions for perfect blindness (i.e. 0-blindness) and $\epsilon$-blind-verifiability discussed in Section \ref{sec:security} for exponentially small $\epsilon$ \cite{dunjko2014composable}. The UBQC protocol can also be used as a method to remotely prepare the states used in the ABE protocol, resulting in a hybrid protocol which requires only single qubit state preparations \cite{kapourniotis2015optimising}. 

\subsection{Measurement \label{sec:measurement}}

\begin{figure}[!h]
\includegraphics[]{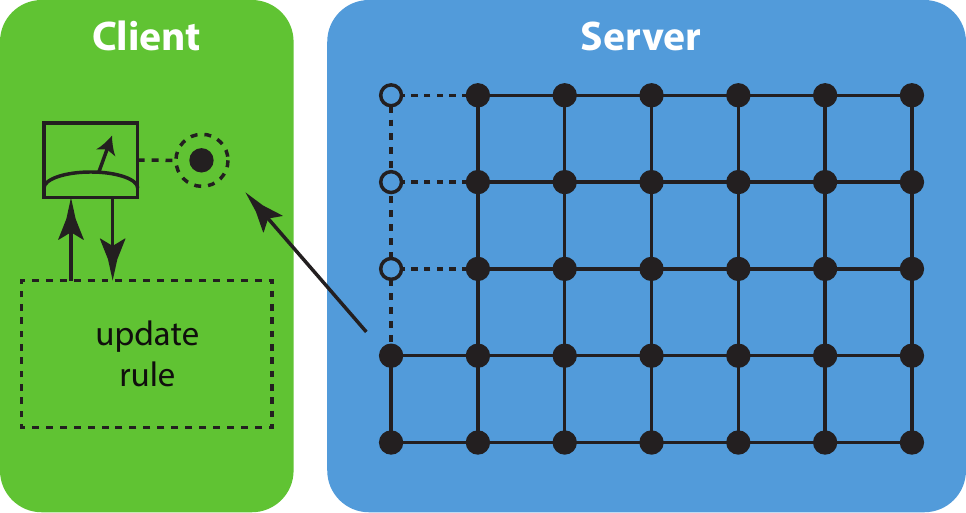}
\caption{A depiction of the BQC approach considered by Morimae and Fujii~\cite{morimae2013blind}, in which the client performs adaptive measurements on a sequence of qubits sent to them by the server. By having the server send the client a universal resource state one qubit at a time, the client can implement an arbitrary computation without ever sending information to the server beyond the initial graph description.\label{fig:AM}}
\end{figure}

An alternate take on the use of MBQC for blind computation was proposed by Morimae and Fujii \cite{morimae2013blind}, which changed the role of the client from state preparation to measurement. They noted that if the client were capable of making the adaptive single-qubit measurements necessary to drive a measurement-based computation on a fixed graph state, then a BQC protocol could be achieved by having the server feed the resource state to the client, one qubit at a time, as depicted in Figure \ref{fig:AM}. Such a scheme is trivially blind, since the direction of communication is always from server to client, and has been shown to satisfy composable security definitions \cite{dunjko2014composable,morimae2013composable}. This approach to blind computation can be thought of as dual to the UBQC approach, due to the fact that in quantum mechanics there is a symmetry under post-selection between preparation and measurement. As a result, fault-tolerance constructions are interchangeable between the two settings, and trap-based verification techniques have been successfully adapted to this setting \cite{morimae2014verification}. Furthermore, the fact that the server is required only to prepare a fixed state has lead to new approaches to verification based on directly testing that this state has been prepared using stabiliser measurements \cite{hayashi2015verifiable,morimae2016measurement}. 

Arguments have been made in both directions over whether state preparation or measurement is a more practical option for the client's device. This will of course depend on the physical implementation under consideration. At least at present, however, photons appear to be the only reasonable choice for long range quantum communication. In this setting, there is indeed a case that measurement may be easier than state preparation due to the difficulty of constructing deterministic single photon sources and the relative ease with which photons can be detected at short wavelengths. There are some gaps in this line of reasoning, however, since it has been shown that weak coherent pulses \cite{dunjko2012blind}, easily producible by an attenuated laser, suffice for the state preparation approach. Furthermore, BQC schemes in which the client makes measurements directly on a resource state prepared by the server are extremely susceptible to photon loss and hence require both extremely high efficiency detectors and near lossless communications links. One possible way to take advantage of the best features of each approach is to make use of an a hybrid of the two. It was noted in \cite{broadbent2009universal} that the classical-quantum (CQ) correlations between the client and server necessary to implement the UBQC protocol could be achieved by measurements made on half of an entangled state with the other half held by the server. This is simply due to the fact that measuring one qubit of a singlet state in a basis $B_\theta$ results in the other qubit being projected onto either $\frac{1}{\sqrt{2}}\left(\ket{0} \pm e^{i\theta}\ket{1}\right)$ where the sign depends on the outcome of the measurement. Thus by having the server feed the client one half of an entangled pair for each qubit to be prepared, the client can effectively remotely prepare the input states used by the server in state-preparation BQC and verification protocols. This can be combined with self-testing of the underlying singlet states to achieve verification in a device-independent fashion \cite{hajduvsek2015device,gheorghiu2015rigidity}. Importantly, in this approach failure by the client to detect a photon is not a significant problem, since the procedure can be repeated until remote preparation of the requisite number of qubits has been achieved, resulting in overhead which scales only inversely with the probability of a single measurement attempt succeeding. Thus a client making use of a single low quality detector and lossy links could still make use of BQC.

\section{BQC with multiple servers} 

\begin{figure}[!h]
\includegraphics[]{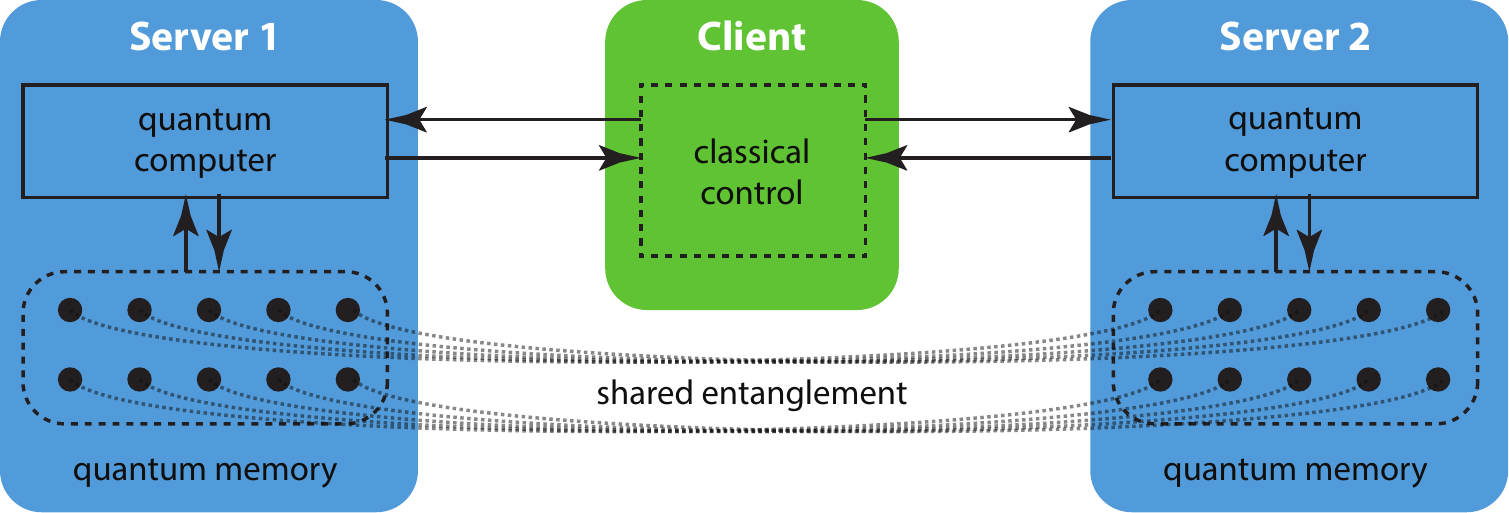}
\caption{The scenario considered by current multi-server protocols. The client communicates classically with two or more servers. The servers are prohibited from communicating with each other directly, but are required to share a large number of entangled qubits.\label{fig:multiserver}}
\end{figure}

The insight discussed in the previous section, that the CQ correlations necessary to implement the UBQC protocol could be achieved through measurements on one half of a bipartite entangled state, immediately gives rise to a version of the UBQC protocol, also introduced in \cite{broadbent2009universal}, in which the client is entirely classical. This required the addition of a second server which shares an entangled state with the first. Importantly, the two servers are restricted from communicating with each other. This situation is depicted in Figure \ref{fig:multiserver}. By requesting the second server measure their half of the shared entangled state in a specified random basis, the client can establish that the correct initial distribution of states with the first server without revealing anything about their computation. They can then proceed with the UBQC protocol as usual. While blindness in this setting follows directly from the proof of blindness in the case where the client prepares and transmits quantum states, the situation for verification is more complicated and the argument for security initially proposed in \cite{broadbent2009universal} does not hold for the most general adversarial behaviour. 

The first schemes to achieve delegation and verification of quantum computation by an entirely classical client to a set of entangled servers were introduced by Reichardt, Unger and Vazirani \cite{reichardt2013classical,reichardt2012classical}  and McKague \cite{mckague2013interactive} using self-testing techniques based on CHSH games. While the structure of the protocols varies significantly, neither disclose the computation to the servers and thus both can be considered blind. The construction considered by Reichardt \textit{et al} has the advantage of requiring only two servers, while that of McKague requires a significantly larger number of servers. One advantage of McKague's protocol over the RUV protocol is that the incurred overhead scales more gently ($O(n^{22})$ as compared to $O(n^{8191})$ where $n$ is the sum of the number of logical gates and qubits). Subsequent efforts were made to lower this overhead by making use of self-tested remote state preparation as input for a verifiable blind computation protocol (that of \cite{fitzsimons2012unconditionally}), proposed independently in somewhat different forms in \cite{hajduvsek2015device} and \cite{gheorghiu2015robustness}. These managed to reduce the the overhead to $O(n^4 \log n)$ in the setting of polynomially many servers \cite{hajduvsek2015device}, a scaling subsequently also achieved by \cite{hayashi2016self} by other means, and $O(n^{2048})$ for the two server setting \cite{gheorghiu2015robustness}. Subsequent work has sought to improve this overhead requirement further \cite{gheorghiu2015rigidity}. In this regard, a new self-testing procedure introduced by Natarajan and Vidick \cite{natarajan2016robust} which does not scale with the number of entangled pairs to be verified shows particular promise.

The above mentioned protocols share at least two common features. Each requires a number of rounds of communication which scales polynomially with depth of the computation and each is either natively blind or can be made blind with trivial adaptations. Recently, however, a class of non-blind verification protocols have emerged based on one-round interactive proofs for the local Hamiltonian which require extremely little communication \cite{fitzsimons2015multiprover,ji2015classical}. These can be exploited to yield verification a method for verifying delegated quantum computation by constructing a Hamiltonian to encode the chosen computation and its purported output and then executing an interactive proof to verify that the ground state energy of this Hamiltonian is below a fixed threshold \cite{fitzsimons2015post,morimae2016post,natarajan2016robust}. This casts into question the link between verification and blindness.

Multi-server approaches have the advantage of eliminating the need for any quantum capability on the part of the client, but this comes at a significant price, both in terms of overhead (particularly for verifiable schemes) and in terms of the additional assumptions regarding a lack of communication between servers. These additional assumptions mean that it is likely impossible in practice to maintain blindness indefinitely, due to the lack of a physical mechanism to prevent communication between servers over arbitrary timescales in an adversarial setting.

\section{Computing on encrypted data and homomorphic encryption}

As noted in Section \ref{sec:security}, blind quantum computation can be achieved by applying a fixed unitary operation to an input which encodes the full circuit to be evaluated together with any associated input states. This forms a concrete link to the notion of computing on encrypted data, a regime similar to blind computation but in which the operation to be evaluated is public. This corresponds to the setting  initially considered by Childs, as discussed in Section \ref{sec:computing}. Progress on protocols of this type has continues to be made, with Fisher \textit{et al} introducing a protocol which reduced the requirements on the client to preparation of specific single qubit states, similar to the requirement for the UBQC protocol, while implementing logic gates in a manner similar to Childs' original scheme \cite{fisher2014quantum}. This approach necessarily sacrificed the native ability to hide the computation being performed, unless a large programmable circuit were to be implemented. However, the result of this sacrifice was a significant saving in communications requirements over the UBQC protocol in the regime where only the data must be hidden, since Clifford group gates could be evaluated without quantum communication. Furthermore, this protocol has been proven secure under an appropriate composable security definition in a companion paper by Broadbent \cite{broadbent2015delegating}. In this context, Broadbent also proposed a mechanism for verifying that a specified computation had been carried out correctly \cite{broadbent2015verify}, formulated as an interactive proof for BQP, making use of hidden computations for which the expected outcome can be computed by the client, similar in spirit to the trap techniques proposed in \cite{broadbent2009universal} and \cite{fitzsimons2012unconditionally}.

Fisher \textit{et al} were far from alone in seeking to reduce the communications overhead required for delegating quantum computation. In the context of blind quantum computation the minimum communications requirements have been analysed both in terms of total communication requirements \cite{giovannetti2013efficient} and in terms of quantum communication requirements \cite{mantri2013optimal} for a number of client settings, with protocols proposed which come close to saturating these lower bounds. Surprisingly, it has been shown that a client with the ability to adaptively prepare multi-qubit states can make use of an iterated teleportation procedure in order to delegate certain quantum computations with exponentially less communication than is required to classically describe the circuit being implemented \cite{perez2015iterated}. 

In the world of classical cryptography, the state of the art regarding computation on encrypted data has changed dramatically over the years since the introduction of the first BQC protocols, with the advent of fully homomorphic encryption \cite{gentry2009fully}. Fully homomorphic encryption schemes allow data to be processed arbitrarily in its encrypted form, without the need for the encryption key. Such encryption schemes allow for computing on encrypted data without need for communication between client and server during the processing. Achieving such a reduction in round complexity in the context of delegated quantum computation would be highly desirable. It should be noted, however, that even in the classical context this has only been achieved by making use of assumptions about the hardness of certain computational problems, such as the approximate shortest vector problem \cite{micciancio2001shortest,khot2005hardness} or the learning with errors problem \cite{regev2009lattices}. While for many cryptographic problems quantum mechanics has offer to replace computational security guarantees with information theoretic ones, this does not appear to be the case here. No-go theorems ruling out evaluation of arbitrary quantum circuits on encrypted data with perfect \cite{yu2014limitations} and near-perfect \cite{newmanquantum} information theoretic security \footnote{It should be noted that the term ``quantum homomorphic encryption'' was first coined by Min Liang \cite{liang2013symmetric}, though it was used to refer to a class of interactive protocols more akin to computing on encrypted data than to a quantum analog of classical homomorphic encryption.}. Nonetheless some progress has been made on developing quantum analogues of homomorphic encryption. Several works have explored the use of partially-homomorphic encryption which support models of computation not classically simulable, including the Boson sampling model \cite{aaronson2011computational}, under weakened information theoretic security guarantees \cite{rohde2012quantum,tan2016quantum}. Broadbent and Jeffery introduced a homomorphic encryption scheme which leveraged a computationally secure classical homomorphic encryption scheme to enable evaluation of circuits containing only a constant number of non-Clifford gates on encrypted quantum data \cite{broadbent2015quantum}. In the case of Broadbent and Jeffery's scheme, the size of the encoding scaled exponentially with the number of non-Clifford gates. However subsequent work by Dulek, Schaffner and Speelman reduced this to only polynomial overhead, resulting in a computationally secure levelled homomorphic encryption scheme \cite{dulek2016quantum}. These latter protocols can be seen as part of a wider programme to broaden quantum cryptographic techniques through the incorporation of computational assumptions \cite{alagic2016computational}. A recent proposal achieves similar functionality to the Broadbent-Jeffery scheme but under an information theoretic security definition, provided a sufficiently large key is used \cite{ouyang2015quantum}. To date, however, no such counterpart to the work of Dulek, Schaffner and Speelman has been found. 

\section{Physical implementations}

The discussion thus far has focused on theoretical constructions for delegated computation protocols. However, if these protocols are to become anything other than theoretical curiosities, it is important to establish a path to physical realisation. To this end, a number of works have sought to make blind and verifiable computation protocols more accessible to experiment. This has taken several forms. While early proposals noted that their constructions could be made fault-tolerant \cite{broadbent2009universal,aharonov2010proceedings}, significant work has since been put into computing explicit fault-tolerance thresholds for blind computation \cite{morimae2012blind,chien2015fault} and dealing with the issue of noise occurring during quantum communication between client and server \cite{chien2015fault,takeuchi2016blind,sheng2016blind}. Issues of fault-tolerance have also been examined in the context of verification \cite{takeuchi2016practically,kashefi2015optimised,fujii2016verifiable}, a subtle topic which is often under-examined in the literature. In the context of multiple-server protocols, the issue of overcoming noisy correlations between servers has also been addressed in the form of entanglement distillation protocols \cite{morimae2013secure,sheng2015deterministic}. Aside from the issue of noise, recent work has also sought enable blind computation based on resource states potentially more amenable to experiment, including proposed BQC protocols based on continuous variables \cite{morimae2012continuous} and the ground states of certain Hamiltonians \cite{morimae2015ground}. In particular, the adaptation of the UBQC protocol to use client-supplied weak coherent states, rather than individual qubits, has received significant attention due to the potential to significantly reduce the technological burden on the client \cite{dunjko2012blind,xu2015blind,li2017blind}.

\begin{figure}[!h]
\includegraphics[]{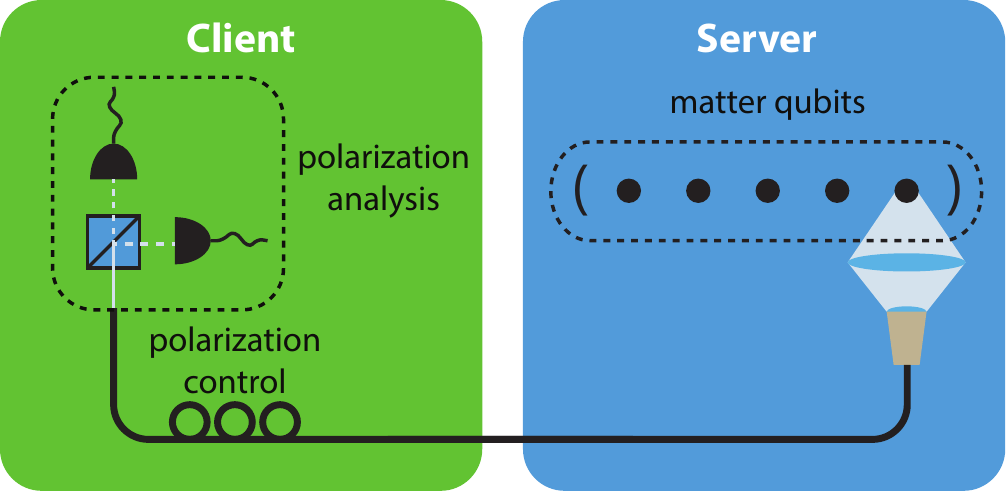}
\caption{Hybrid architectures combining matter qubits contained in the server with optical qubits for communication present a possible path to overcoming some of the hurdles faced by the current generation of experiments. One potential approach is to make use of trapped ions or similar matter qubits to perform the computation which can emit photons entangled with the matter system. Using the remote state preparation approach described in Section \ref{sec:measurement}, measurements made by the client could then be used to remotely establish correlations with the matter qubits in the server sufficient to implement UBQC-like protocols. In the example depicted above the state of the matter qubits is taken to be entangled with polarization degree of the photon, however other degrees of freedom could also be used. The ability to create such entangled states between matter and photonic qubits has already been demonstrated for a number of systems \cite{blinov2004observation,togan2010quantum}. \label{fig:ions}}
\end{figure}

To date, several delegated computation protocols have been successfully demonstrated in a quantum optics setting. Barz \textit{et al} performed a successful demonstration of the UBQC protocol using a four qubit resource state entangled according to a variety of graphs, corresponding to subgraphs of the brickwork state \cite{barz2012demonstration}. The reported experiments blindly implemented both the Deutsch-Josza algorithm \cite{deutsch1992rapid} and Grover's search algorithm \cite{grover1997quantum}, although in each case experimental limitations required that only certain measurement were hidden, corresponding to the choice of oracle for each problem. Subsequently, similar experiments were used to demonstration of a verifiable blind computation protocol modified from theoretical work previously presented in \cite{fitzsimons2012unconditionally}. The client-measuring protocol of Morimae and Fujii has also been demonstrated using a four qubit cluster state \cite{greganti2016demonstration}, which included a replication of the verification method used in \cite{barz2013experimental} in the new client setting. Finally, in the paper proposing their protocol for computing on encrypted data, Fisher \textit{et al} included an experimental demonstration using photonic qubits, where they demonstrated individual quantum gates sufficient for universal quantum computation \cite{fisher2014quantum}. 

While photons are the clear choice for communication between client and server in blind computation protocols, they are less well suited to the business of performing large scale computation. This is due to a combination of factors including the destructive nature of current measurement techniques and the difficulty of performing deterministic entangling gates between photonic qubits. Thus, while universal computation is possible with linear optics \cite{briegel2009measurement}, a hybrid system, such as that depicted in Figure \ref{fig:ions}, involving matter qubits for computation and photonic qubits for communication may prove a more viable path for scalable blind quantum computation.

\section{Conclusions and outlook}

Given recent developments in quantum technologies and the longstanding paradigm in classical computing of delegating computationally intensive tasks to shared systems, the emerging interest in delegated quantum computation is both understandable and timely. While the progress discussed in this review illustrates the potential of blind quantum computation and related protocols, the field is still in its infancy, with new results coming on a regular basis but with many open questions still remaining. Perhaps the most prominent open question is that of whether or not blind or verifiable computation is possible with a single server and a completely classical client. In this setting, even when multiple non-entangled non-communicating servers are allowed, the existence of secure protocols for blind and verifiable computation remains an open question. Indeed, the precise relationship between blindness and verification is currently unresolved. In the context of homomorphic encryption, the existence of fully homomorphic quantum encryption under plausible computational assumptions remains open, despite the promising progress of Dulek \textit{et al} \cite{dulek2016quantum}. In the context of verification, the most significant challenges facing the field include the necessity to drastically reduce overhead and sensitivity to noise of current device-independent verification protocols, and the development of methods to verify analogue quantum simulators and other special purpose devices. While some progress has been made on the question of verifying non-universal devices \cite{kapourniotis2014blindness,hangleiter2016direct,carolan2014experimental}, much more progress in this direction is necessary to fully unlock the potential of such devices. Lastly it should be noted that progress to date has only scratched the surface of the variety of functionality that future quantum networks may unlock. Recent developments, in terms of multi-user blind computation \cite{kashefi2016blind} and publicly verifiable quantum computation \cite{honda2016publicly}, together with established results on secure multi-party quantum computation \cite{crepeau2002secure,ben2006secure,colbeck2009quantum} give some indication of the potential for new secure quantum computing protocols beyond the two party setting. Given these open questions, there is the potential for significant theoretical advance in the coming years. Harnessing the latest advances in experimental capabilities to go beyond the current generation of proof-of-principle experiments is also likely to be an important future direction.

\section*{Acknowledgements}

The author thanks Vedran Dunjko, Atul Mantri,  Yingkai Ouyang, Christopher Portmann and Renato Renner for helpful comments on the manuscript. The author acknowledges support from the Air Force Office of Scientific Research under AOARD grant FA2386-15-1-4082 and the Singapore National Research Foundation under NRF Award NRF-NRFF2013-01.

\bibliographystyle{apsrev}
\bibliography{review}

\end{document}